\documentclass[conference]{IEEEtran}
\IEEEoverridecommandlockouts
\usepackage{cite}
\usepackage{amsmath,amssymb,amsfonts}
\usepackage{algorithmic}
\usepackage{tikz}
\usepackage{graphicx}
\usepackage{textcomp}
\usepackage{xcolor}
\def\BibTeX{{\rm B\kern-.05em{\sc i\kern-.025em b}\kern-.08em
    T\kern-.1667em\lower.7ex\hbox{E}\kern-.125emX}}
\begin{document}

\newsavebox{\fundinglogo}

\sbox{\fundinglogo}{%
    \begin{tikzpicture}[baseline, overlay]
        \node[anchor=center] at (6em,-6pt) {\includegraphics[width = 0.4\columnwidth]{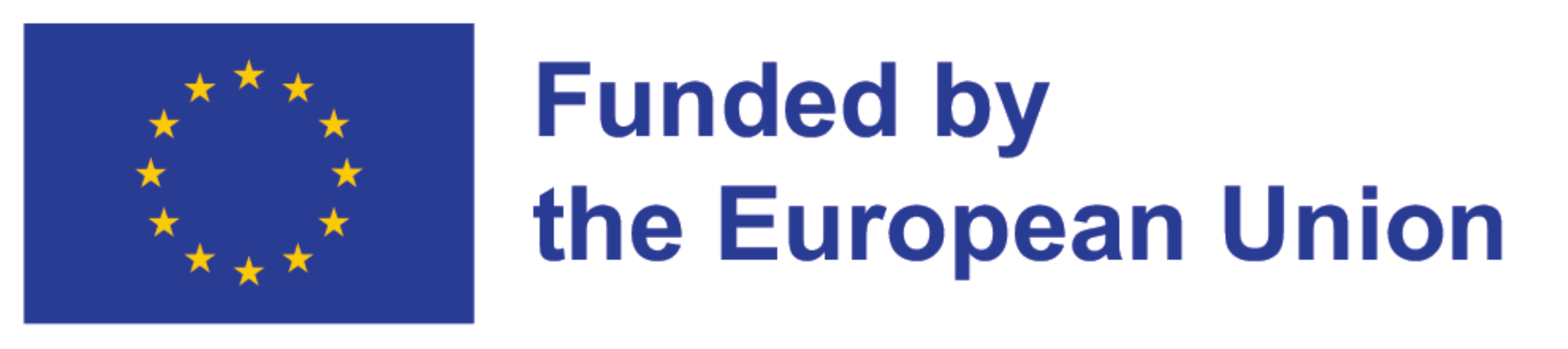}}; 
    \end{tikzpicture}%
}

\title{Eigenmode analysis of a half-mode uniplanar metamaterial-inspired substrate integrated waveguide
\thanks{
Copyright (C) 2026 IEEE. Personal use of this material is permitted.  Permission from IEEE must be obtained for all other uses, in any current or future media, including reprinting/republishing this material for advertising or promotional purposes, creating new collective works, for resale or redistribution to servers or lists, or reuse of any copyrighted component of this work in other works.\\
This project has received funding from the European
Union's Horizon 2020 research and innovation programme
under the Marie Sklodowska-Curie grant agreement
No.101146306. 
\protect\usebox{\fundinglogo}
}
}

\author{
\IEEEauthorblockN{Maria-Thaleia Passia}
\IEEEauthorblockA{\textit{School of Electrical}\\ 
{and Computer Engineering} \\
\textit{Aristotle University of Thessaloniki}\\
Thessaloniki, Greece\\
passiamg@ece.auth.gr}
\and
\IEEEauthorblockN{Kyriakos Katsaros-Gkouskos}
\IEEEauthorblockA{\textit{School of Electrical}\\ 
{and Computer Engineering} \\
\textit{Aristotle University of Thessaloniki}\\
Thessaloniki, Greece \\
kgouskos@hotmail.com
}
\and
\IEEEauthorblockN{Traianos V. Yioultsis}
\IEEEauthorblockA{\textit{School of Electrical}\\ 
{and Computer Engineering} \\
\textit{Aristotle University of Thessaloniki}\\
Thessaloniki, Greece \\
traianos@ece.auth.gr}
}

\IEEEaftertitletext{\vspace{-2\baselineskip}}

\maketitle

\begin{abstract}
In this work, we systematically analyze the propagation characteristics of a new half-mode uniplanar substrate integrated waveguide (SIW) based on complementary split-ring resonators (CSRR), using a finite element method (FEM) eigenmode solver. The proposed half-mode CSRR SIW has a simpler fabrication than the SIW, since the via are substituted by CSRRs, and is more compact than the existing full uniplanar CSRR SIW, since its transverse size is reduced almost by half.  To gain insight into the propagation characteristics of the proposed half-mode CSRR SIW and guide its synthesis process, we solve an eigenvalue problem that determines the complex propagation constant of the supported modes. By calculating the dispersion diagrams of the dominant mode, with all loss mechanisms included, we demonstrate that the half-mode uniplanar CSRR SIW retains the performance of the existing full CSRR SIW. 
\end{abstract}


\section{Introduction}

The SIW is a low-loss planar transmission line, based on metalized via holes, that has dominated microwave and mmWave communications~\cite{b1}. As communications shift toward higher frequencies, the via dimensions are considerably reduced, rendering drilling holes and injecting metalization a rather complex and costly fabrication process.

To avoid the fabrication complexity of metalized via holes, several via-less SIW variations have been proposed that can be implemented using standard PCB manufacturing. The corrugated SIW (CSIW) uses quarter-wave open-circuit stubs instead of via to form the electric sidewalls~\cite{b2}. A different solution is the use of metamaterial-inspired SIWs, where each via row is replaced by a metasurface (MS), capable of restricting in-plane propagation~\cite{b3}. Such SIWs may potentially offer greater design flexibility, as a different number of rows or even a non-uniform MS may be selected. Among various metamaterial-inspired SIWs, the uniplanar single-CSRR SIW is the preferred choice, as it offers the simplest fabrication by using grounded single CSRRs to form the electric walls~\cite{b3}, while maintaining performance (Fig.~\ref{fig:CSRR}(a)). However, it uses two rows of CSRRs to substitute each via row, which leads to a greater footprint than the SIW. Hence, it would be beneficial to synthesize a more compact single-CSRR SIW variation, to facilitate component integration.

In this work, we introduce the half-mode uniplanar single-CSRR SIW, which occupies nearly half the footprint of the uniplanar single-CSRR SIW, while maintaining comparable performance (Fig.~\ref{fig:CSRR}(b)). Half-mode SIWs based on the SIW and CSIW have already been synthesized~\cite{b4,b5,b6}.
To gain insight into the propagation characteristics of this new transmission line and determine appropriate geometric parameters that minimize transmission losses, we employ an eigenmode solver based on the FEM. The $\beta-\omega$ formulation is most commonly used, especially by commercial FEM solvers; however, it only calculates the dispersion diagram in terms of the propagation constant $\beta$. Accurately calculating the attenuation constant as well is necessary to compare the half-mode CSRR SIW's performance to that of existing SIWs. Hence, we employ an $\omega-k$ formulation, which can calculate both the propagation and attenuation constant~\cite{b7,b8}. We incorporate all loss mechanisms into the formulation and implement it in the weak form of COMSOL Multiphysics. We determine suitable parameters for the half-mode CSRR SIW and compare its propagation characteristics with the full variation.

\section{Dispersion diagrams of the half-mode single-CSRR SIW}
\subsection{Formulation}
To determine the propagation characteristics of the half-mode single-CSRR SIW we use a full wave 3D eigenmode solver~\cite{b7}.  The formulation employs a curl-curl Helmholtz equation and is applied on the waveguide's unit cell. The direction of propagation $\mathbf{\hat{k}}$ and the angular frequency $\omega$ are assumed known, whereas the complex wavenumber $k = \beta - j \alpha$, with $\beta$ being the propagation and $\alpha$ the attenuation constant, is considered the unknown eigenvalue of the problem. To impose Floquet periodic boundary conditions on the master and slave planes, the following periodic field transformation is applied $
    \mathbf{E} = \mathbf{e} e^{-jk\mathbf{\hat{k} \cdot r}}$.
\begin{figure}
	\centering
		\includegraphics[width=0.48\columnwidth,trim=0cm 0cm 0cm 0cm,clip]{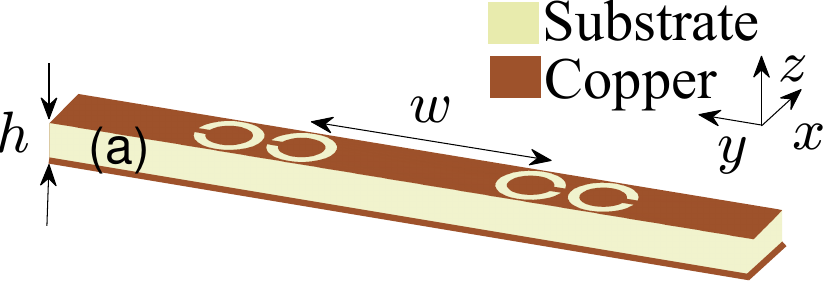}
	\includegraphics[width=0.48\columnwidth,trim=0cm 0cm 0cm 0cm,clip]{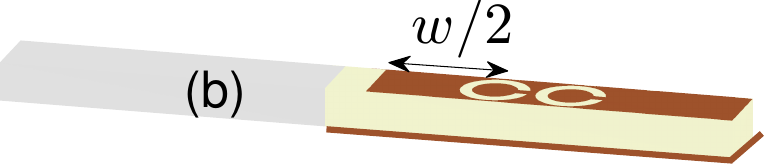} \\
 		\includegraphics[width=0.45\columnwidth,trim=0cm 0cm 0cm 0cm,clip]{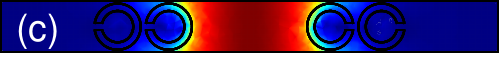}
        \includegraphics[width=0.49\columnwidth,trim=0cm 0cm 0cm 0cm,clip]{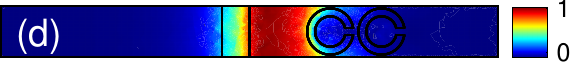}   
  	\caption{Configuration and electric field distribution $|E_z|$ of the dominant mode for the uniplanar single-CSRR SIW  (a,c) and  half-mode uniplanar single-CSRR SIW unit cells (b,d).}
	  \label{fig:CSRR}
\end{figure}
By applying this transformation, the Floquet periodic boundary conditions are replaced by simpler continuity conditions for the transformed fields. The curl-curl Helmholtz equation is written in terms of the transformed field. A quadratic eigenvalue problem (QEP) is formed, with $k$ being the eigenvalue and $\mathbf{e}$ the eigenvector. The QEP is transformed into a generalized eigenvalue problem of double size which is in the following form:
\begin{equation}
    \begin{bmatrix}
       \begin{bmatrix}
          \mathbf{C}
       \end{bmatrix}   &  \begin{bmatrix}\mathbf{B}
      \end{bmatrix}  \\
           \begin{bmatrix}
          \mathbf{0}
      \end{bmatrix}   &  \begin{bmatrix}
          \mathbf{I}
      \end{bmatrix} 
    \end{bmatrix} \begin{bmatrix}
        \mathbf{e} \\ k \mathbf{e}
    \end{bmatrix} = k  \begin{bmatrix}
       \begin{bmatrix}
          \mathbf{0}
       \end{bmatrix}   &  \begin{bmatrix}\mathbf{-A}
      \end{bmatrix}  \\
           \begin{bmatrix}
          \mathbf{I}
      \end{bmatrix}   &  \begin{bmatrix}
          \mathbf{0}
      \end{bmatrix} 
    \end{bmatrix}  \begin{bmatrix}
        \mathbf{e} \\ k \mathbf{e}
    \end{bmatrix} 
\end{equation}
We analyze the half-mode uniplanar CSRR SIW unit cell, which has 1D periodicity along the $x$-axis (Fig.~\ref{fig:CSRR}(b)). We impose impedance boundary conditions (IBCs) on metal surfaces to account for conductor losses.  The IBC is written in terms of the electric field envelopes as:
\begin{equation} \label{eq:IBC1}
    \mathbf{\hat{n}_{\rm ext}} \times (\mathbf{e} e^{-jk\mathbf{\hat{k} \cdot r}}) = Z_s (\mathbf{h} e^{-jk\mathbf{\hat{k} \cdot r}}),
\end{equation}
where $Z_s = (1+j)R_s$,  $R_s = \sqrt{ \frac{\omega \mu} {2 \sigma_c}}$ and $\sigma_c=5.8\times 10^7$ S/m is copper's conductivity. Eq.~\ref{eq:IBC1} is transformed into the following equation to form the surface term:
\begin{equation} \label{eq:IBC2}
    \mathbf{\hat{n}_{\rm ext}} \times \mu_r^{-1}(\nabla \times \mathbf{e} -j k \mathbf{\hat{x}} \times \mathbf{e}) = \frac{j \omega \mu_0}{Z_s} \mathbf{e_t},
\end{equation} 
Eq.(\ref{eq:IBC2}) is introduced into the surface integral to model the metal surfaces. Dielectric losses are also included in the formulation by assuming complex permittivity values.

\subsection{Results}
We calculate the dispersion diagrams  and  field distributions of the half-mode  uniplanar single-CSRR SIW's dominant mode. 
We design the half-mode single-CSRR SIW so that low losses appear around 26 GHz. A substrate of $\varepsilon_r~=~2.18~(1~-~j~0.0009)$ and thickness $h$=1.05156~mm are used.  By performing a parametric analysis of the geometric parameters, we determine those which result in lower total losses: external CSRR radius $r$=0.93~mm, CSRR ring width $c$~=~0.32~mm and CSRR gap $g$=0.3~mm and $w$=6.5~mm. 

Similarly to the SIW, the dominant mode of the uniplanar single-CSRR SIW is TE$_{10}$, with the maximum electric field located at the center of the waveguide (Fig.~\ref{fig:CSRR}(c)). Hence,  the center plane can be considered as an equivalent magnetic wall. By dividing the uniplanar single-CSRR SIW into two parts, each can support a half of the field distribution independently due to the large width-to-height ratio~\cite{b5}. Hence, the dominant mode of the half-mode variation is the TE$_{0.5,0}$ (Fig.~\ref{fig:CSRR}(d)). As shown in Fig.~\ref{fig:CSRR}(b), the grounded substrate is slightly extended beyond the center to enclose any fringing fields. 

\begin{figure}
	\centering
\includegraphics[width=1\columnwidth,trim=0cm 0cm 0cm 0cm,clip]{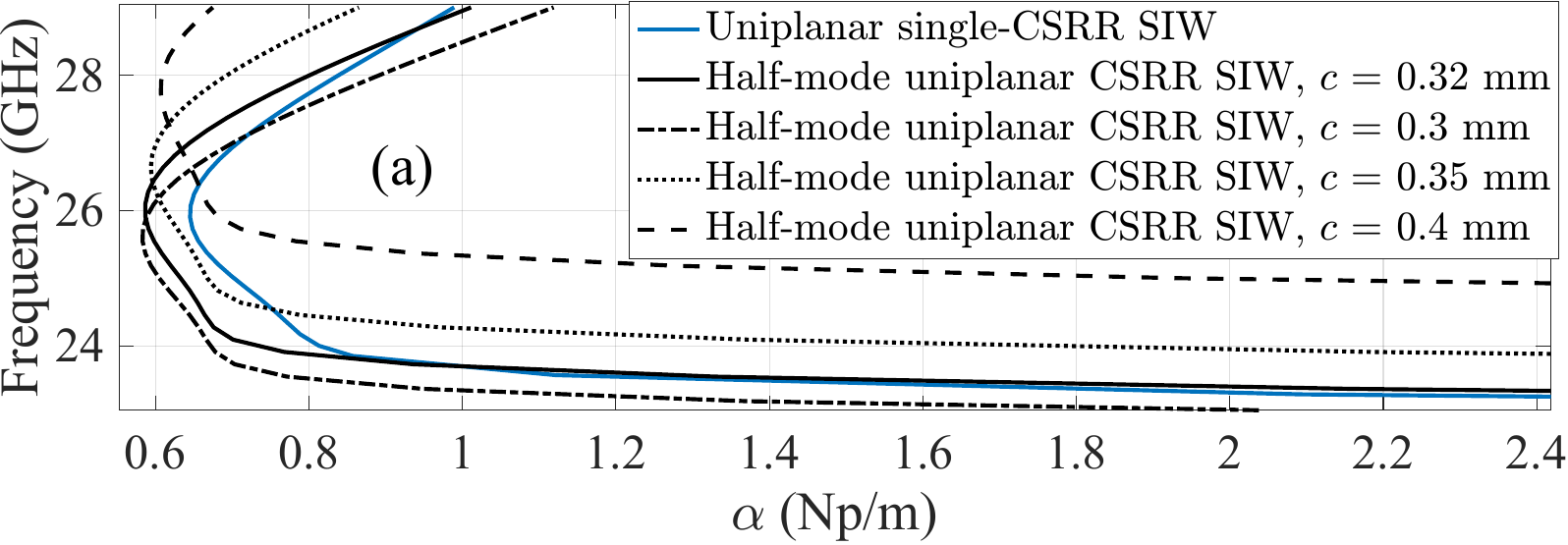}
\includegraphics[width=1\columnwidth,trim=0cm 0cm 0cm 0cm,clip]{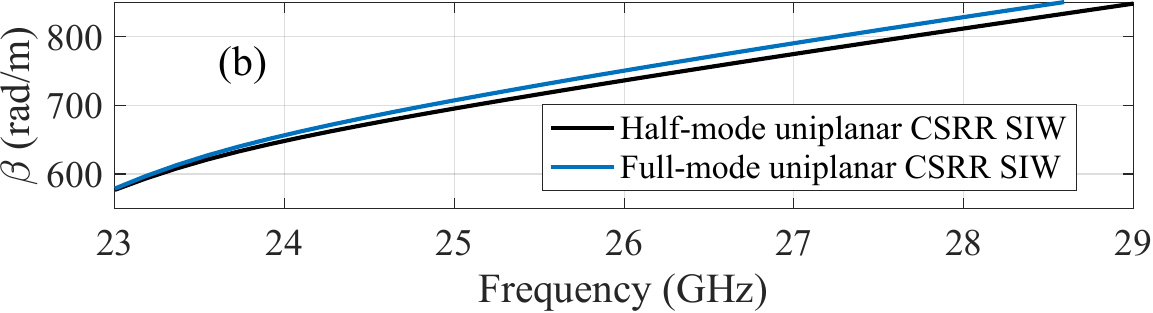}        
  	\caption{The attenuation constant of the half-mode single-CSRR SIW, varying  $c$. A comparison to the full single-CSRR SIW is also included.}
	  \label{fig:a_w}
\end{figure}

The dispersion diagrams of the dominant mode of the half-mode and full single-CSRR SIWs are shown in Fig.~\ref{fig:a_w} in terms of the attenuation (a) and propagation (b) constant. All loss mechanisms, dielectric, conductor, and radiation losses, are considered. The full uniplanar single CSRR SIW has the dimensions stated above, but with $c$~=~0.3~mm, to place the lower-loss range around 26 GHz. For the half-mode single-CSRR SIW we show the effect of changing the ring width $c$. The lower-loss range shifts toward higher frequencies as the ring width increases. We observe that the half-mode variation has similar or slightly lower losses and a similar propagation constant to the full variation, hence, confirming its merit as a low-loss compact alternative.

\section{Conclusion}
The half-mode uniplanar CSRR SIW has similar or slightly lower losses than the  uniplanar CSRR SIW and offers a more compact alternative for mmWave applications.

\end{document}